\documentclass[preprint2,twocolappendix,trackchanges]{aastex6}

\usepackage{booktabs}
\usepackage{bm}

\begin{document}


\title{Extended relativistic configuration interaction and many-body perturbation calculations of Spectroscopic Data for the $n \leq 6$ Configurations\\ in Ne-like Ions Between Cr XV and Kr~XXVII}

\author{K. Wang\altaffilmark{1,2,3}, Z.B. Chen\altaffilmark{4}, R. Si\altaffilmark{3}, P. J\"{o}nsson\altaffilmark{5},  J. Ekman\altaffilmark{5}, X.L. Guo\altaffilmark{6,3}, S. Li\altaffilmark{2}, F.Y. Long\altaffilmark{2}, W. Dang\altaffilmark{1}, X.H. Zhao\altaffilmark{1}, 
\\R. Hutton\altaffilmark{3}, C.Y. Chen\altaffilmark{3}, J. Yan\altaffilmark{2,7,8}, and X. Yang\altaffilmark{9}}

\email{chychen@fudan.edu.cn}
\email{yan$_$jun@iapcm.ac.cn}

\affil{$^1$Hebei Key Lab of Optic-electronic Information and Materials, The College of Physics Science and Technology, Hebei University, Baoding 071002, China\\
	$^2$Institute of Applied Physics and Computational Mathematics, Beijing 100088, China\\
    $^3$Shanghai EBIT Lab, Institute of Modern Physics, Department of Nuclear Science and Technology, Fudan University, Shanghai 200433, China\\
	$^4$College of Science,  National University of Defense Technology, Changsha 410073, China\\
   $^5$Group for Materials Science and Applied Mathematics, Malm\"o University, SE-20506, Malm\"o, Sweden\\
   	$^6$Department of Radiotherapy, Shanghai Changhai Hospital, Second Military Medical University, Shanghai 200433, China\\
	$^7$Center for Applied Physics and Technology, Peking University, Beijing 100871, China\\
	$^8$Collaborative Innovation Center of IFSA (CICIFSA), Shanghai Jiao Tong University, Shanghai 200240, China\\
	$^9$The Third Institute of Surveying and Mapping of Hebei Province, Hebei Bureau of Geoinformation, Shijiazhuang 050000, China}




\begin{abstract}
Level energies, wavelengths, electric dipole, magnetic dipole, electric quadrupole, and magnetic quadrupole transition rates, oscillator strengths, and line strengths from combined relativistic configuration interaction and many-body perturbation calculations are reported for the 201 fine-structure states of the $2s^2 2p^6$, $2s^2 2p^5 3l$, $2s 2p^6 3l$, $2s^2 2p^5 4l$, $2s 2p^6 4l$, $2s^2 2p^5 5l$, and $2s^2 2p^5 6l$ configurations in all Ne-like ions between \ion{Cr}{15} and \ion{Kr}{27}. 
Calculated level energies and transition data are compared with experiments from the NIST and CHIANTI databases, and other recent benchmark calculations. The mean energy difference with the NIST experiments is only 0.05\%. 
The present calculations significantly increase the amount of accurate spectroscopic data for the $n >3$ states in a number of Ne-like ions of astrophysics interest. A complete dataset should be helpful in analyzing new observations from the solar and other astrophysical sources, and is also likely to be useful for modeling and diagnosing a variety of plasmas including astronomical and fusion plasma.
\end{abstract}



\section{INTRODUCTION}\label{sect:in}
The rapid advance of astronomical observations requires more extensive accurate spectroscopic data. This paper is a continuation of our recent work of providing the data of energy levels and transition characteristics for L-shell ions to the accuracy needed to exploit the high quality of observations from space- and ground- based telescopes. Systematic calculations for the beryllium, carbon and nitrogen isoelectronic sequences have already been performed~\citep{Wang.2014.V215.p26,Wang.2015.V218.p16,Wang.2016.V223.p3}. In this paper, we report accurate data for the neon isoelectronic sequence between \ion{Cr}{15} and \ion{Kr}{27}.

In view of a stable closed L-shell ground state, Ne-like ions show high abundance over a wide range of temperatures in ionization equilibrium~\citep{Mazzotta.1998.V133.p403,Bryans.2006.V167.p343,Bryans.2009.V691.p1540,Liang.2010.V518.p64}. A wealth of emission lines in a wide wavelength range are frequently observed in astrophysics~\citep{Feldman.2000.V544.p508,Behar.2001.V548.p966,Mewe.2001.V368.p888,Kaastra.2002.V386.p427,Ko.2002.V578.p979,Raassen.2002.V389.p228,Ness.2003.V598.p1277,Curdt.2004.V427.p1045,Holczer.2005.V632.p788,Landi.2005.V160.p286,Brown.2008.V176.p511,DelZanna.2008.V481.p69,Shestov.2008.V34.p33,Warren.2008.V685.p1277,Raassen.2013.V550.p55,DelZanna.2014.V565.p14,Shestov.2014.V780.p15}. These observations constitute an important tool for obtaining useful information of the physical conditions, chemical abundances, and evolution of the astrophysical objects. For example, in high-resolution observations with the \emph{Chandra} and \emph{XMM-Newton} X-ray observatories, the \ion{Fe}{17} spectrum dominated the X-ray emission in the 700-1000 eV range of a large number of astrophysical objects. Thus these spectral lines were used for  diagnostics~\citep{Paerels.2003.V41.p291,DelZanna.2011.V536.p59}. The \ion{Fe}{17} EUV lines were measured by the \emph{Hinode} Imaging Spectrometer and provided useful information about the nature of the heating in the solar corona~\citep{Culhane.2007.V243.p19,DelZanna.2009.V508.p1517}. The \ion{Ni}{19} lines have been identified in the spectra of solar flares~\citep{Phillips.1982.V256.p774,Landi.2005.V160.p286}, the Capella~\citep{Behar.2001.V548.p966}, and the supergiant star~\citep{Raassen.2013.V550.p55}, and offer an opportunity for determining elemental abundances and physical conditions of astrophysical objects.

 Using various methods a number of  calculations have been carried out to provide datasets of energy structures and transition rates for the Ne-like sequence~\citep{Cogordan.1985.V31.p545,Quinet.1991.V44.p164,Hibbert.1993.V53.p23,Dong.2003.V146.p161,Dong.2003.V205.p87,Fischer.2004.V87.p1,Gu.2005.V156.p105,Ishikawa.2009.V79.p25301,DelZanna.2009.V508.p1517,Jonsson.2014.V100.p1}. However, in these studies the calculations were restricted to the $n \leq 3$ states (the 37 fine-structure states of the $(1s^2) 2s^2 2p^6$, $2s^2 2p^5 3l$, and $2s 2p^6 3l$ configurations). 

Atomic data involving higher-lying states of the $n>3$ configurations are also urgently demanded because of their wide applications for line identifications and plasma diagnostics in solar physics and astrophysics ~\citep{Phillips.1982.V256.p774,Acton.1985.V291.p865,DelZanna.2008.V481.p69,DelZanna.2009.V508.p1517,Raassen.2013.V550.p55,DelZanna.2014.V565.p14}. 
Calculations were preformed for the $n > 3$ states in \ion{Fe}{17} using various methods, including the calculations of ~\citet{Chen.2003.V36.p453} and~\cite{Nahar.2003.V408.p789} using the configuration interaction (CI) method of the code SUPERSTRUCTURE~\citep{Eissner.1974.V8.p270}, and the calculation by~\citet{Aggarwal.2004.V420.p783} utilizing the GRASP code of~\cite{Dyall.1989.V55.p425}.
Relativistic perturbation theory with a model potential was used to calculate transitions probabilities of the lowest 72 excited energy to the ground state for ions up to $Z=66$~\citep{Ivanova.1991.V49.p1}. 
Using mixed CI and perturbation theory, energies and oscillator strengths for the seven lowest $J=1$ odd excited states of neon-like ions with $Z=11-18$ were calculated by~\citet{Savukov.2003.V36.p4789}. 
Relativistic combined configuration interaction (RCI) and many-body perturbation theory  (MBPT) calculations were carried out for  wavelengths of $n\rightarrow2$ ($3\leq n\leq7$) transitions in \ion{Fe}{17} and \ion{Ni}{19}~\citep{Gu.2007.V169.p154}. 
~\citet{Liang.2010.V518.p64} reported the results for the energy levels, and transition data among the 209 states of the $2s^2 2p^6$, $(2s, 2p)^7 nl$ ($n \leq 5$ and $l \leq n-1 $), and $2s^2 2p^5 n^\prime l^\prime$ ($6 \leq n^\prime \leq 7$ and $l^\prime \leq 2 $) configurations in Ne-like ions from \ion{Na}{2} to \ion{Kr}{27} using the AUTOSTRUCTURE code~\citep{Badnell.1986.V19.p3827}. Among the above $n > 3$ calculations, the MBPT results of~\citet{Gu.2007.V169.p154} in \ion{Fe}{17} and \ion{Ni}{19} are sufficiently accurate to identify observed spectra. In this work, however, transition properties were not computed. The other mentioned calculations are not adequate to meet the accuracy requirements of line identification and interpretation in astrophysics. 

The present work aims at extending the accurate calculations for \ion{Fe}{17} and \ion{Ni}{19} by ~\citet{Gu.2007.V169.p154}, providing the energy data of spectroscopic accuracy and transition rates for the $n \leq 6$ states in a number of Ne-like ions of astrophysics interest. 
By using a combined RCI and MBPT approach in the FAC code~\citep{Gu.2003.V582.p1241,Gu.2005.V89.p267,Gu.2005.V156.p105,Gu.2006.V641.p1227}, we present data for the lowest 201 bound energy states arising from the $2s^2 2p^6$, $(2s, 2p)^7 nl$ ($3 \leq n \leq 4$ and $l \leq n-1$), and  $2s^2 2p^5 n^\prime l^\prime$ ($5 \leq n^\prime \leq 6$ and $l^\prime \leq n^\prime -1$) configurations in Ne-like ions from \ion{Cr}{15} to \ion{Kr}{27}, as well as the electric-dipole (E1), electric-quadrupole (E2), magnetic-dipole (M1), and magnetic-quadrupole (M2) transition rates among these states. 
To assess the accuracy of the MBPT data,  the multiconfiguration Dirac-Hartree-Fock (MCDHF) and RCI method has been used to calculate the data for \ion{Fe}{17} (hereafter referred to as MCDHF/RCI). The MBPT energies in \ion{Fe}{17} agree well with the MCDHF/RCI values, as well as the experimental energies from the Atomic Spectra Database (ASD) of the National Institute of Standards and Technology (NIST)~\citep{Kramida.2015.V.p}. The energy differences between the calculated MBPT and  MCDHF/RCI level energies are within 0.07\% for all 201 states in \ion{Fe}{17}, and the mean difference of the NIST and MBPT values is 0.05\% for the 425 states listed in the NIST ASD.
Compared with the recent systematic MCDHF and RCI calculations by~\citet{Jonsson.2014.V100.p1}, in which both accurate energy levels and transition rates were given, the present calculations are extended to report the data for additional 174  levels of the $2s 2p^6 3l$, $2s^2 2p^5 4l$, $2s 2p^6 4l$, $2s^2 2p^5 5l$, and $2s^2 2p^5 6l$ configurations. The calculations also extend the elaborate work by \citet{Gu.2005.V156.p105,Gu.2007.V169.p154} to include data of additional eleven neon-like ions between \ion{Cr}{15} and \ion{Kr}{27}. The excellent description of the energy separations along the sequence makes it possible to point out a number of lines for which the experimental identifications can be questioned. A complete dataset including energy levels and transition data should be helpful in analyzing new data from the solar and other astrophysical sources.

\section{Theory}
\subsection{The MBPT method}
According to the Rayleigh-Schr\"{o}dinger perturbation theory, the no--pair Dirac--Coulomb--Breit
(DCB) Hamiltonian $H_{\mathrm{DCB}}$ for an $N$-electron ionic system can be written as~\citep{Sucher.1980.V22.p348,Gu.2005.V89.p267,Gu.2005.V156.p105}:
\begin{equation}
H_{\mathrm{DCB}}=\sum_{i}^{N}[h_{d}(i)-\frac{Z}{r_{i}}]+\sum_{i<j}^{N}(\frac{1}{r_{ij}}+B_{ij}),
\end{equation}
where $h_{d}(i)$ and $Z$ are the free-electron Dirac Hamiltonian and the nuclear charge, respectively. $r_{i}$ and $r_{ij}$ are the radial coordinate of  electron $i$, and the distance between the electrons $i$ and $j$, respectively. $B_{ij}$ is the frequency independent Breit interaction, given by
\begin{equation}
B_{ij}=-\frac{1}{2r_{ij}}[\bm{\alpha}_{i}\cdot\bm{\alpha}_{j}+\frac{(\bm{\alpha}_{i}\cdot \bm{r}_{ij})(\bm{\alpha}_{j}\cdot \bm{r}_{ij})}{r_{ij}^{2}}].
\end{equation}
where $\bm \alpha_{i}$ is a matrix vector constructed from Pauli spin matrices. $H_{\mathrm{DCB}}$ is divided into two parts, namely,  a model Hamiltonian $H_{0}$ and a perturbation $V$, given by
\begin{equation}
H_{0}=\sum_{i}[h_{d}(i)+U(r_{i})],
\end{equation}
\begin{equation}
V=-\sum_{i}[\frac{Z}{r_{i}}+U(r_{i})]+\sum_{i<j}(\frac{1}{r_{ij}}+B_{ij}),
\end{equation}
where $U(r)$ is a model potential including the screening effects of
all electrons, whose appropriate choice makes $V$ as small as possible.

For calculations:

(a).~The approximated local central potential $U(r)$ and eigenfunctions $\Phi_{k}$ of $H_{0}$ are obtained by the Dirac--Fock--Slater self--consistent field calculations.

(b).~The Hilbert space of the Hamiltonian is divided into two parts, namely, a model space $M$, and  the orthogonal space $O$. A subset of $\Phi_{k}$ will define the space $M$, and the remaining states belong to the space $O$.

(c).~The second order eigenvalues are obtained through solving the generalized eigenvalue problem for the first-order effective Hamiltonian.

\subsection{The MCDHF method}
The MCDHF method was described in detail
by~\cite{Grant.2007.V.p}, and here we just give a brief outline. The atomic state function (ASF) is given as an expansion over configuration state functions (CSFs)
\begin{equation}
\Psi(\gamma J\pi)=\sum_j c_j\Phi(\gamma_j J\pi). \label{csftoasf}
\end{equation}
where \emph{J} and $\pi$ are the total angular momentum and parity of the system, respectively, $\gamma_j $ is a set of quantum numbers, additional to $J\pi$, to specify a CSF, and $c_j$ is the mixing coefficient. 

For calculations:

(a).~A CSF $\Phi(\gamma_j J\pi)$ is constructed from a product of
single-electron wave functions through a proper angular momentum coupling and antisymmetrization.

(b).~The self--consistent iteration method is used to obtain simultaneously the Dirac orbitals and the expansion coefficients.

(c).~When the radial orbitals are obtained, RCI calculations are performed, which include the Breit interaction and first-order Quantum Electrodynamics (QED) corrections (self-energy and vacuum polarization).

\section{Calculations and Results}
In the MBPT calculations, the model space \emph{M} contains the configurations $2s^2 2p^6$, $(2s, 2p)^7 nl$ ($3 \leq n \leq 4$ and $l \leq n-1$), and  $2s^2 2p^5 n^\prime l^\prime$ ($5 \leq n^\prime \leq 6$ and $l^\prime \leq n^\prime -1$). The \emph{N} space contains all configurations formed by single and double (SD) virtual excitations of the \emph{M} space. For single/double excitations, configurations with $n\leq 200$ and $l\leq \min{(n-1,25)}$/the inner electron promotion up to $n = 65$ and promotion of the outer electron up to $n^\prime = 200$ are considered. For level energy and radiative transition calculations, some corrections such as finite nuclear size, nuclear recoil, and QED are also included. A more detailed description of the MBPT calculations procedure could be found in our recent work~\citep{Wang.2014.V215.p26,Wang.2015.V218.p16,Wang.2016.V223.p3}.

Table~\ref{tab.lev.sub} displays the computed excitation energies of 201 fine structure levels in Ne-like ions ($Z=24-36$) obtained from the MBPT method. 
Also listed in the table are the experimental energy levels recommended by the NIST ASD. Among the 2613 energy levels in the 13 ions given by the MBPT method, 443 experimental results are available. The wavelengths ($\lambda_{ji}$ in $\rm {\AA}$), line strengths ($S_{ji}$ in atomic units, 1 AU = $\rm 6.460\times 10^{-36} cm^2 esu^2$), weighted oscillator strengths ($gf_{ji}$ dimensionless) and radiative rates ($A_{ji}$ in s$^{-1}$) for the E1, M1, E2, and M2 transitions among the 201 levels for each ion, are listed in Table~\ref{tab.tr.sub}.

For assessing the accuracy of the MBPT results,  the MCDHF and subsequent RCI calculations are carried out for \ion{Fe}{17}. Separate calculations are performed for the even and odd states belonging to the \emph{M} space of the above MBPT calculations, which are considered as the multi-reference configurations. The CSFs  expansions are obtained through single and double excitations of the orbitals in the multi-reference configurations with orbitals in an active set with principal quantum numbers $n = 3,...,8$ and angular symmetries $s$, $p$, $d$, $f$, $g$, $h$, and $i$. To monitor the convergence of the calculated energies and transition parameters, the active sets were increased in a systematic way by adding layers of orbitals. For the $n = 8$ expansion this resulted in 3034729 CSFs with even parity and 3009779 CSFs with odd parity. The self-consistent field calculations for each layer of orbitals are followed by RCI calculations. A more detailed description of the MCDHF/RCI calculations procedure could be found in our recent work~\citep{Jonsson.2013.V184.p2197,Jonsson.2014.V100.p1,Si.2015.V48.p175004,Si.2015.V163.p7}.

\section{Evaluation of data}\label{sect:com}

\subsection{Energy Levels}\label{sect:en}
 Up to now, with regard to experimental data and elaborate computed results along the isoelectronic sequence, the \ion{Fe}{17} spectrum is the most studied in astrophysics. For example, many \ion{Fe}{17} EUV lines observed by the Hinode EUV Imaging Spectrometer were identified by~\citet{DelZanna.2009.V508.p1517}. These \ion{Fe}{17} lines provide useful information about the nature of the heating in the solar corona. In Table~\ref{tab.lev.Fe}, the MBPT energy results for the 201 levels in \ion{Fe}{19} are compared with experimental values of ~\citet{DelZanna.2009.V508.p1517}, who reviewed the \ion{Fe}{17} spectrum in the 30-450 \AA~ range, and provided accurate results for the $n=3-5$ states, which have been included in the CHIANTI database~\citep{DelZanna.2015.V582.p56,Dere.1997.V125.p149}. The present MCDHF/RCI values, the previous results for the $2s^2 2p^6$ and $2s^2 2p^5 3l$ levels~\citep[MCDHF/RCI2]{Jonsson.2014.V100.p1} and the relativistic multireference M\"{o}ller--Plesset results for the $2s^2 2p^6$ and $2l^7 3l^\prime$ states~\citep[MR-MP]{Ishikawa.2009.V79.p25301}, as well as the experimental values from the NIST ASD, are also given in the table for comparison. 
Compared with the present MBPT calculations, ~\citet{Gu.2005.V156.p105} adopted the same method, and reported similar results which are not shown in this table.
 
Compared with the previous elaborate computed results (MCHDF/RCI2 and MR-MP) for the $n=3$ levels, the present MBPT and MCDHF/RCI calculations give very consistent results. 
The experimental values from the NIST and CHIANTI databases and the four theoretical datasets also show good agreement (within 0.1\%) for the $n=3$ states, except for the $2s 2p^6 3s~^1\!S_{0}$ state. 
For this level, the NIST value 869.1 eV is observed at a considerably higher energy (about 4 eV) than the CHIANTI experimental value 865.266 eV and the MBPT, MCHDF/RCI and MR-MP theoretical values (864.8332, 865.2301 and 865.146 eV). 
 
Observed energies are scarce and the identification of some states becomes questionable for the $n > 3$ states. 
The $2s^2 2p^5 4d~^1\!D_{2}$ (1010.682 eV), and $2s^2 2p^5 4f~^1\!G_{4}$ (1017.9 eV) and $2s^2 2p^5 4f~^3\!G_{4}$ (1014.2 eV) states in the NIST ASD do not have any obvious counterparts in the Chianti database or in calculated energies, and misidentification can not be ruled out. 
As an example, we analyze the $2s^2 2p^5 4f~^1\!G_{4}$ (1017.9 eV) state in more detail. By means of the $2s^2 2p^5 3d~^3\!D_3$ level energy, the observed wavelength 58.98~\AA ($2s^2 2p^5 3d~^3\!D_3-2s^2 2p^5 4f~^1\!G_4$) is utilized to extract the $2s^2 2p^5 4f~^1\!G_4$ level energy~\citep{Shirai.2000.V.p}. 
This NIST wavelength is about 1.4\% lower than the CHIANTI, MBPT, and MCDHF/RCI values (59.776, 59.821, and 58.856~\AA), but is very close to the CHIANTI, MBPT, and MCDHF/RCI values (58.980, 58.026, and 59.057~\AA) for the $2s^2 2p^5 3d~^3\!F_3-2s^2 2p^5 4f~^1\!G_4$ transition, whose the lower state is $2s^2 2p^5 3d~^3\!F_3$, but not $2s^2 2p^5 3d~^3\!D_3$. 
And the transition rate is $1.075\times 10^{12}$ s$^{-1}$ for the $2s^2 2p^5 3d~^3\!F_3-2s^2 2p^5 4f~^1\!G_4$ ($\Delta L =1$) transition, which is indeed larger by over one order of magnitude than the $8.8\times 10^{10}$ s$^{-1}$ for the  $2s^2 2p^5 3d~^3\!D_3-2s^2 2p^5 4f~^1\!G_4$ ($\Delta L = 2$) transition. 
Therefore, we conclude that the $\Delta L =1$ transition is more likely to be observed than the  $\Delta L =2$ transition, and the NIST wavelength 58.98 \AA should be assigned to the $2s^2 2p^5 3d~^3\!F_3-2s^2 2p^5 4f~^1\!G_4$ transition. By means of this wavelength and the NIST energy 805.0331 eV of  the $2s^2 2p^5 3d~^3\!F_3$ states, the NIST value for $2s^2 2p^5 4f~^1\!G_4$ should be changed to 1015.3 eV, which agrees with the CHIANTI, MBPT, and MCDHF/RCI (1015.96, 1015.255, and 1015.461 eV) to within 0.1\%.  Based on the above argument, a misidentification for this NIST level cannot be ruled out. Together with the $2s^2 2p^5 4f~^1\!G_{4}$ (1017.9 eV) energy, all the other NIST values for the $2s 2p^6 3s~^1\!S_{0}$, $2s^2 2p^5 4d~^1\!D_{2}$, and $2s^2 2p^5 4f~^3\!G_{4}$ states in \ion{Fe}{17}, for which the NIST results differ from the MBPT values by more than 0.2\%, are tabulated in Table~\ref{tab.lev.nistmbptlardif}. 

The agreement of the CHIANTI experimental energies and the MBPT results is better. Deviations are less than 0.2\% for all 30 $n=4,5$ states listed in the CHIANTI database, and are within 0.1\% for 28 states. We can also see from Table~\ref{tab.lev.Fe} that the present MBPT and MCDHF/RCI calculations give very consistent results for all the 201 $n \leq 6$ levels, and the deviation of the two datasets are within 0.07\% for all levels. The calculations predict energy levels with such a high precision that the results can be utilized to analyze the new observations from space- and ground- based telescopes.
 
\begin{figure*}[h]
	\epsscale{1.05}
	\plotone{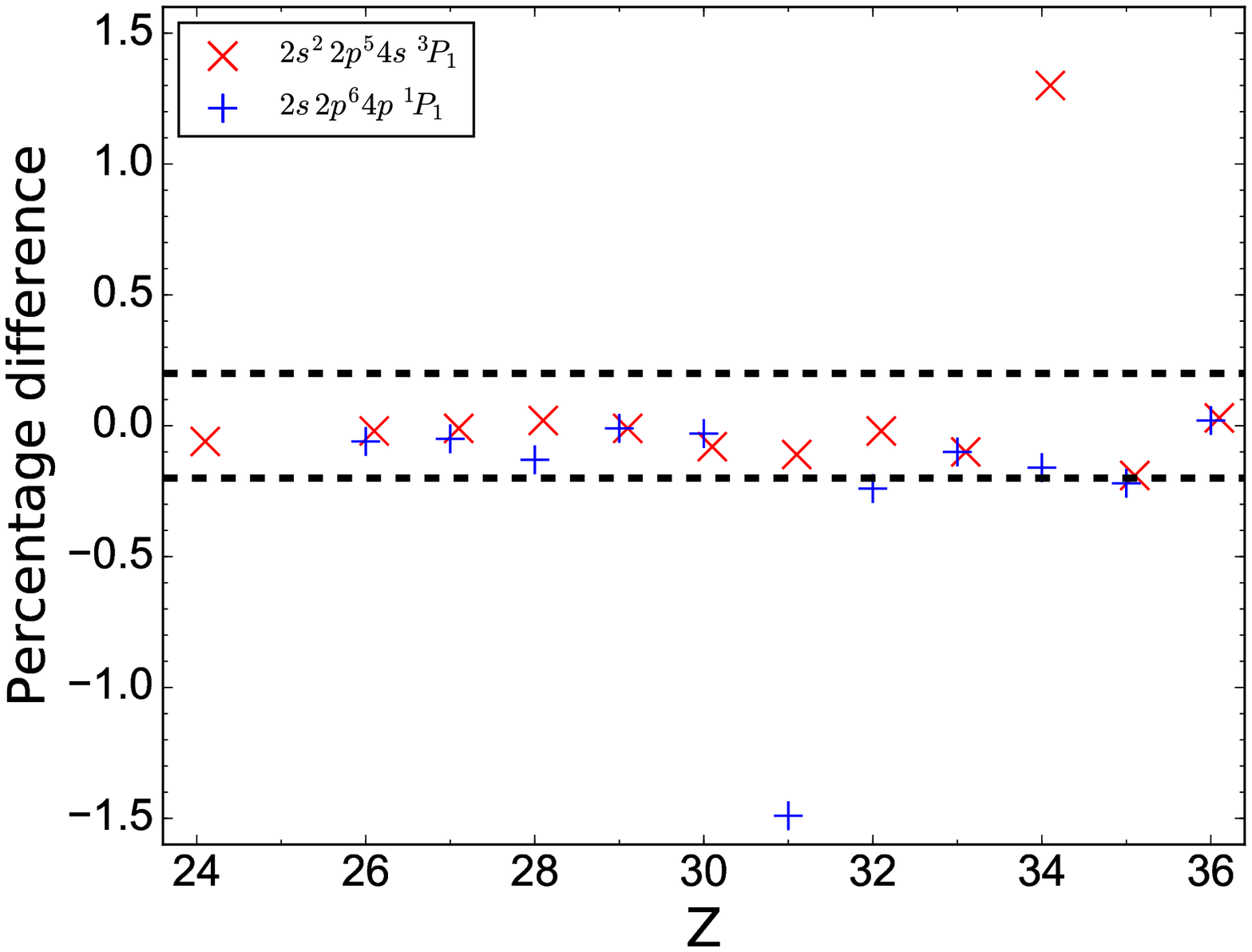}
	\caption{Percentage differences of the MBPT energies relative to the NIST observations for the $2s^2 2p^5 4s\/ ~{}^3\!P_{1}$ and $2s 2p^64p\/ ~{}^1\!P_{1}$ states along the sequence. \label{fig.lev.nistlargedifferences}}
\end{figure*}

For further assessing the accuracy of the MBPT energies, we compare them with the NIST experimental values for all the 13 Ne-like ions.  Among the 2613 energy levels in 13 ions given by the MBPT method, the 443 NIST results are available. The computed energies agree very well with the NIST values. The differences between experimental and calculated energies are less than 0.1\% for 393 states, and are within 0.2\% for another 32 states. 
The remaining 18 states including four levels in \ion{Fe}{17} discussed in detail above, for which the deviations are larger than 0.2\%, are listed in Table~\ref{tab.lev.nistmbptlardif}. We cannot find any obvious duplicate energies in the present MBPT calculations, and these NIST values should be carefully used. As an example, 
Figure~\ref{fig.lev.nistlargedifferences} shows the energy deviations as functions of $Z$ for the $2s^2 2p^5 4s~^3P_1$ and $2s 2p^6 4p~^1P_1$ states. Some obvious anomalies are seen for the $2s^2 2p^5 4s~^3P_1$ state in \ion{Se}{25} (the differene is about 1.3\%), and  the $2s 2p^6 4p~^1P_1$ state (1.5\%) in \ion{Ga}{22}. The differences fall between 0.2\%-0.3\% for the $2s 2p^6 4p~^1P_1$ state in \ion{Ge}{23} and \ion{Br}{26}. The misidentification, line blending,  or large experimental errors of the spectral observations could be responsible for the large uncertainty of the data compiled by the NIST ASD~\citep{Kramida.2015.V.p}. Apart from these irregularities, the two datasets agree well for most states along the sequence.

In short, apart from the 18 states included in Table~\ref{tab.lev.nistmbptlardif}, the mean energy deviation of the observed and computed values for the 425 states included in the NIST ASD is 0.05\%. Seeing that the same computational procedure is adopted for each ion, which implies that the quality of the data should be consistent and systematic, we conclude that relatively large uncertainties of observed energies brings on the large deviations for these states, and these NIST values should be re-evaluated.

\subsection{Radiative Rates}
\begin{figure*}[h]
	\epsscale{1.05}
	\plotone{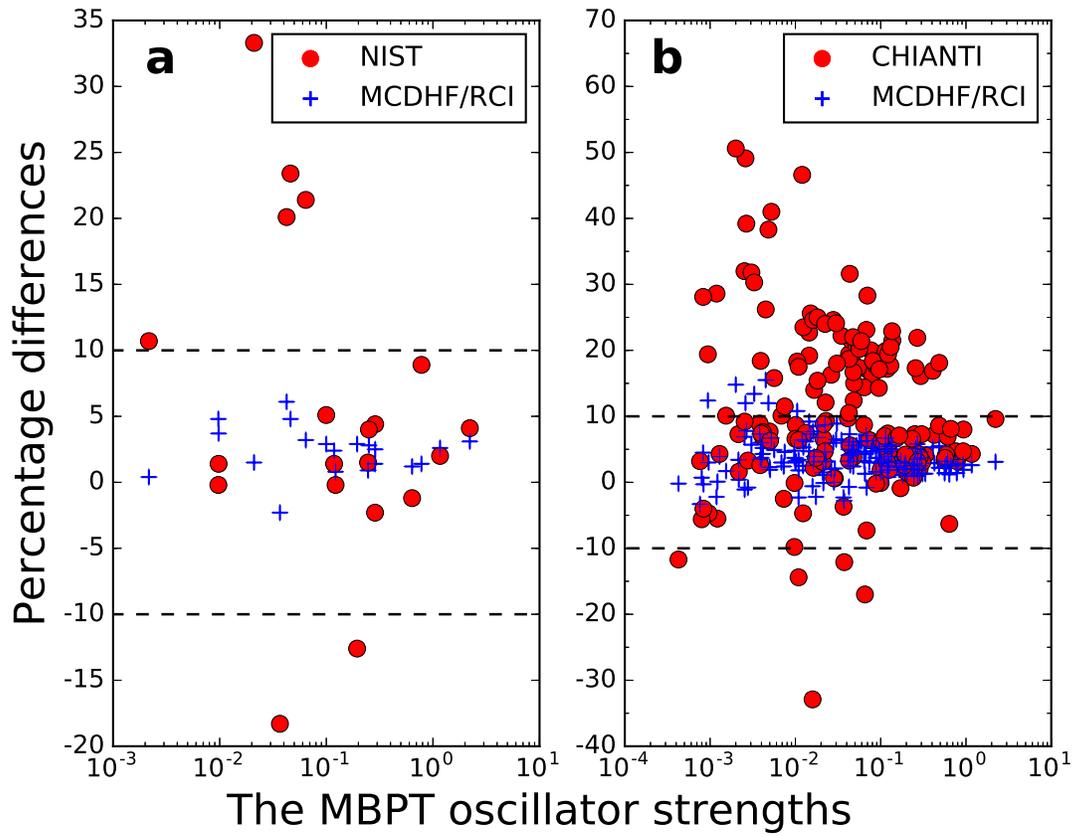}
	\caption{(a) Percentage differences of the NIST and MCDHF/RCI oscillator strengths relative to the present MBPT results for the transitions among the $n \leq 3$ states given by the NIST ASD. (b) Percentage differences of the CHIANTI and MCDHF/RCI oscillator strengths relative to the present MBPT results for the transitions among the $n \leq 3$ states given by the CHIANTI database. Dashed lines indicate the differences of $\pm10\%$. \label{fig.tr.n2.n3.nist.chianti}}
\end{figure*}

In Table~\ref{tab.tr.n2.n3}, weighted oscillator strengths for the E1, M1, E2, and M2 transitions among the $n \leq 3$ levels of the $2s^2 2p^6$, $2s^2 2p^5 3s$, $3p$, and $3d$, and $2s 2p^6 3s$, $3p$ and $3d$ configurations are shown. Our results, $gf$(MBPT) and $gf$(MCDHF/RCI), are compared for \ion{Fe}{17} with the calculated values from~\citet{Jonsson.2014.V100.p1}, $gf$(MCDHF/RCI2), and the NIST ASD~\citep{Kramida.2015.V.p}, $gf$(NIST). The overall agreement among the present MBPT and MCDHF/RCI values  and the previous MCDHF/RC2 results is good, and the relative deviations are within 10\% for the most of transitions. The average differences (with standard deviations) are $3.1\% \pm 4.4\%$ between the MBPT and MCDHF/RCI values and $2.1\% \pm 3.1\%$ between the MBPT and MCDHF/RCI2 values, which are also satisfactory. Among the large number of the transitions listed in Table~\ref{tab.tr.n2.n3}, the $gf$ values for some transitions (20 transitions) are given by the NIST ASD. The NIST $gf$ values for these 20 transitions are compared with the MBPT $gf$ values in Figure~\ref{fig.tr.n2.n3.nist.chianti}~(a). The two datasets agree within 10\% for 13 transitions, while differing from each other between 10\% and 35\% for the 7 transitions. Note that good agreement (within 6\%) can be found between the MBPT and MCDHF/RCI $gf$ values for all transitions in Figure~\ref{fig.tr.n2.n3.nist.chianti}~(a), and thus the NIST values for these 7 transitions, which are compiled by~\citet{Fuhr.1988.V17(Supplement4).p1}, should be updated.  

Weighted oscillator strengths among the $n \leq 3 $ states in \ion{Fe}{17} given by the CHIANTI database are also compared with the present MBPT $gf$ values  in Figure~\ref{fig.tr.n2.n3.nist.chianti}~(b). Many of the CHIANTI compilations differ from the present calculations by 10\%-50\%, and moreover, the deviations exceed 10\% for many relatively strong transitions with $gf$ values $\geq 10^{-2}$. The agreement of the present two calculations is within 10\%  for all strong transitions, and is no more than 15\% for few weak transitions.  

\begin{figure*}[h]
	\epsscale{1.05}
	\plotone{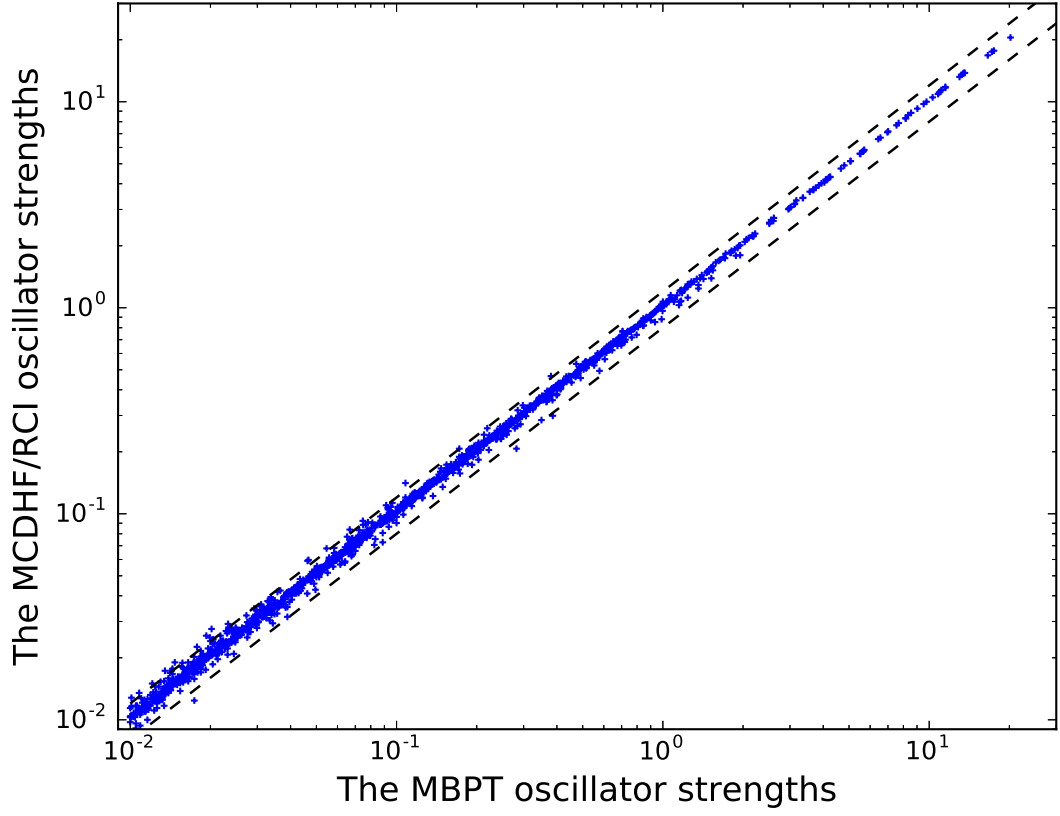}
	\caption{Comparison of the present MBPT oscillator strengths with the MCDHF/RCI results for the transitions with $gf \geq 0.01$ in \ion{Fe}{17}. Dashed lines indicate the differences of $\pm 20\%$. \label{fig.tr.Fe.gf.mbpt.mcdf}}
\end{figure*}

\begin{figure*}[h]
	\epsscale{1.05}
	\plotone{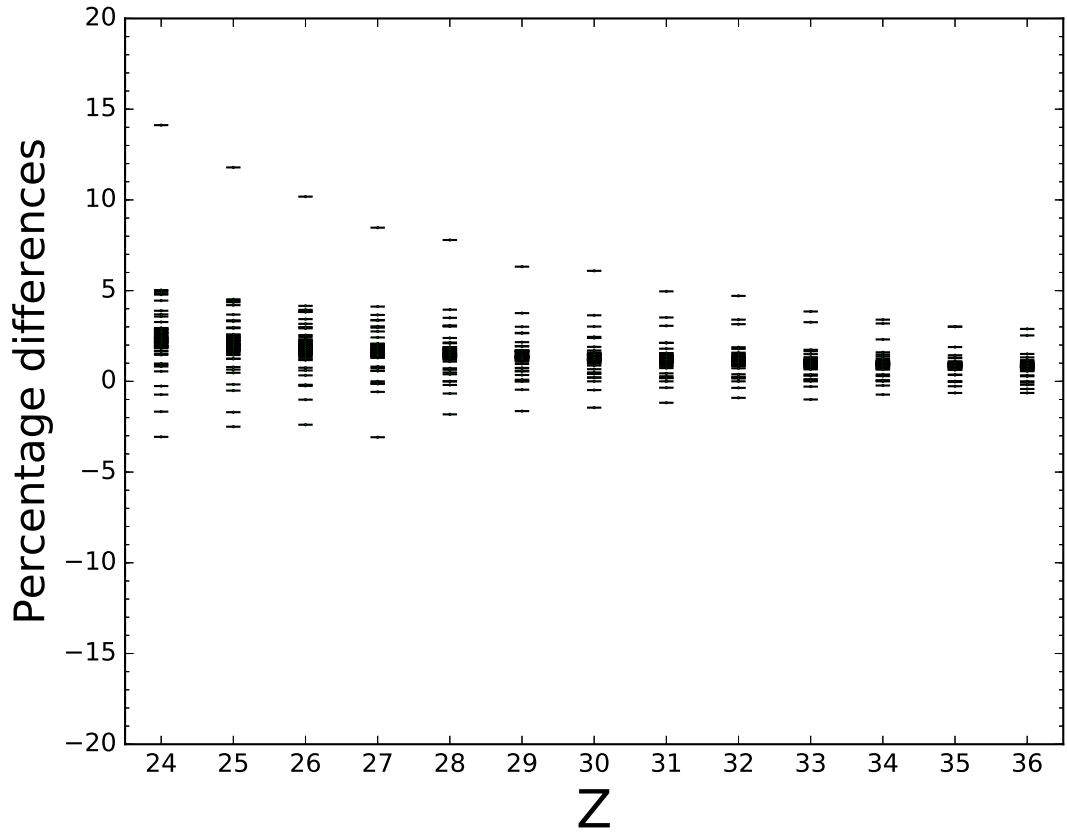}
	\caption{Comparison between the MCDHF/RCI2 and MBPT oscillator strengths for the $gf \geq 0.01$ transitions among the $n \leq 3$ states  along the sequence. \label{fig.tr.n2.n3.gf.mbpt.mcdf2}}
\end{figure*}

To further assess the accuracy of the present calculations, in Figure~\ref{fig.tr.Fe.gf.mbpt.mcdf} the MCDHF/RCI weighted oscillator strengths are compared with the MBPT values for all the 1557 strong transitions ($gf \geq 10^{-2}$) among the $n \leq 6$ states in \ion{Fe}{17}, and the comparison of the MBPT and MCDHF/RCI2 calculations for all the 675 strong transitions among the $n \leq 3$ states from \ion{Cr}{15} to \ion{Kr}{27} are shown in Figure~\ref{fig.tr.n2.n3.gf.mbpt.mcdf2}.  For 92\% of the transitions in \ion{Fe}{17} shown in Figure~\ref{fig.tr.Fe.gf.mbpt.mcdf}, the agreement of the present two calculations are within 10\%, while they differ from each other by over 20\% (but less than 40\%) for only 29 transitions. The upper states of these 29 transitions mostly belong to the highest states of the $n=6$ configurations. For such transitions, the present MCDHF/RCI calculations converge very slowly with increasing active sets. Nevertheless, the average difference with the standard deviation of the present two calculations for the 1557 transitions is only 3.0\%$\pm$ 5.5\%. In addition, as shown in Figure~\ref{fig.tr.n2.n3.gf.mbpt.mcdf2} the MBPT and MCDHF/RCI2 $gf$ values for the 675 transitions among the $n \leq 3$ states from \ion{Cr}{15} to \ion{Kr}{27}  agree within 10\% for 672 transitions. The average difference with the standard deviation of the two calculations for all transitions is only 1.4\%$\pm$ 1.2\%, which is highly satisfactory. 

Based on the above analysis we conclude that the present transitions data have better accuracy compared to the values from the NIST and CHIANTI databases. Using the part of transition values with insufficient accuracy, especially for the strong transitions, may lead to quite different, even wrong results when carrying out line identifications and plasma diagnostics in solar physics and astrophysics. Therefore, hopefully, it would be possible to replace the existing CHIANTI data, as well as the NIST values by the present MBPT and/or MCDHF/RCI results. 

\begin{figure*}[h]
	\epsscale{1.05}
	\plotone{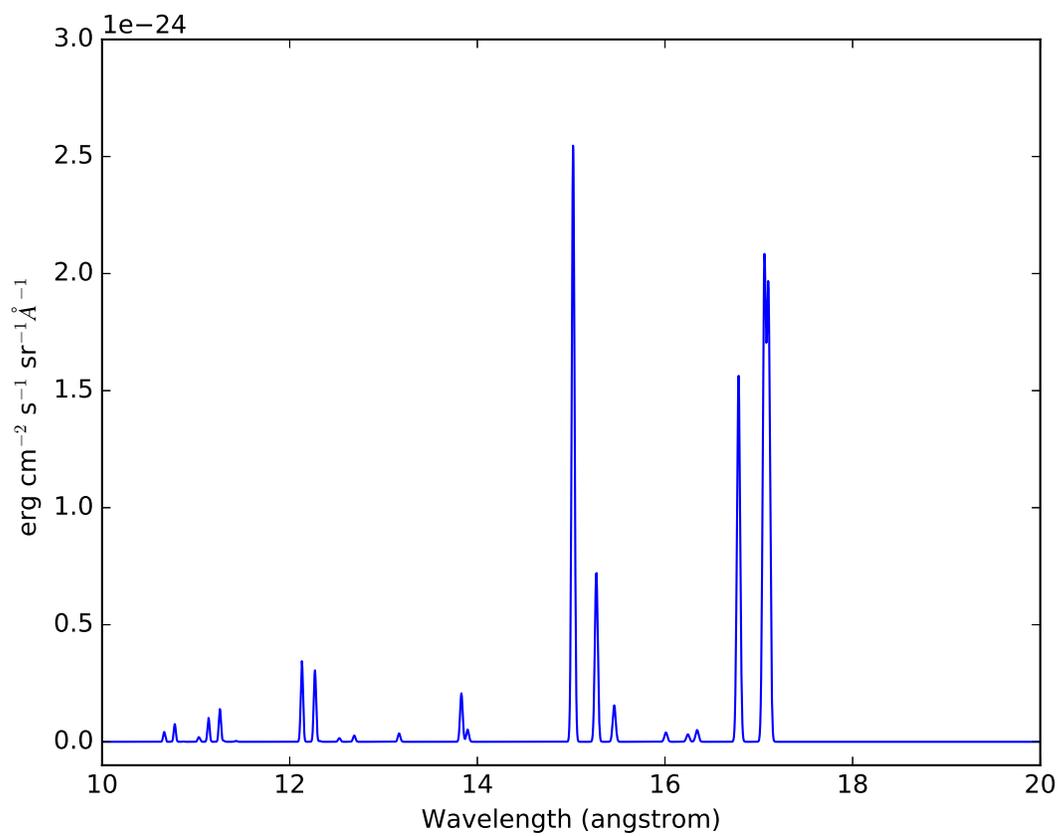}
	\caption{Synthetic Fe XXI spectrum containing transitions between 10~--~20 \AA. See text for details. \label{fig.spectrum}}
\end{figure*}

~\citet{DelZanna.2011.V536.p59} have pointed out that the \ion{Fe}{17} lines in the X-rays range can be reliably used for the measurement of electron temperatures in the solar corona and other astrophysical sources. 
Using the MBPT radiative transition data, as well as the collisional atomic data recommended by the CHIANTI database,  in conjunction with the statistical equilibrium code of Dufton~\citep{Dufton.1977.V13.p25}, the synthetic \ion{Fe}{17} spectra in the range of 10~--~20 \AA~are shown in Figure~\ref{fig.spectrum}. The intensity of each transition is represented by a Gaussian distribution with a resolving power of 1000, corresponding to a temperature $T_e = 10^7$ K and a density $N_e = 10^{11}$ cm$^{-3}$ , a typical solar flare condition. As shown in Figure~\ref{fig.spectrum}, prominent transitions (with wavelengths and transition rates) in the 10~--~20 \AA~range are 

\vspace{5mm}
{\small $2s^2 2p^6~^1\!S_0- 2s^2 2p^5 5d~^1\!P_1$ (11.256 \AA~and $3.07\times10^{12}$~s$^{-1}$)
\vspace{1.0mm}
	
	$2s^2 2p^6~^1\!S_0- 2s^2 2p^5 4d~^1\!P_1$ (12.130 \AA~and $5.62\times10^{12}$~s$^{-1}$)
\vspace{1.0mm}

	$2s^2 2p^6~^1\!S_0- 2s^2 2p^5 4d~^3\!D_1$ (12.269 \AA~and $5.08\times10^{12}$~s$^{-1}$)
\vspace{1.0mm}
	
	$2s^2 2p^6~^1\!S_0- 2s 2p^6 3p~^1\!P_1$ (13.830 \AA~and $3.33\times10^{12}$~s$^{-1}$)
\vspace{1.0mm}
	
	$2s^2 2p^6~^1\!S_0- 2s^2 2p^5 3d~^3\!D_1$ (15.268 \AA~and $6.08\times10^{12}$~s$^{-1}$)
\vspace{1.0mm}
	
	$2s^2 2p^6~^1\!S_0- 2s^2 2p^5 3d~^3\!P_1$ (15.459 \AA~and $9.04\times10^{10}$~s$^{-1}$)
\vspace{1.0mm}
	
	$2s^2 2p^6~^1\!S_0- 2s^2 2p^5 3s~^3\!P_1$ (16.784 \AA~and $7.88\times10^{11}$~s$^{-1}$)
\vspace{1.0mm}
	
	$2s^2 2p^6~^1\!S_0- 2s^2 2p^5 3s~^1\!P_1$ (17.059 \AA~and $9.34\times10^{11}$~s$^{-1}$)
\vspace{1.0mm}
	
	$2s^2 2p^6~^1\!S_0- 2s^2 2p^5 3s~^3\!P_2$~(17.103 \AA~and $2.06\times10^{5}$~s$^{-1}$)
	}
	
\vspace{5mm}
The strongest resonance transition in the spectrum is

\vspace{5mm}
{\small	$2s^2 2p^6~^1\!S_0- 2s^2 2p^5 3d~^1\!P_1$ (15.021 \AA~and $2.19\times10^{13}$~s$^{-1}$)
	\vspace{1.0mm}}



\section{SUMMARY}
Systematic and consistent MBPT calculations have been preformed in Ne-like ions with $Z=24-36$ using the FAC code. A complete dataset with high accuracy, including energies, wavelengths, line strengths, oscillator strengths, and transition rates for the E1, M1, E2, and M2 transitions among the 201 states of the $2s^2 2p^6$, $(2s, 2p)^7 3l$, $(2s, 2p)^7 4l$, $2s^2 2p^5 5l$, and $2s^2 2p^5 6l$ configurations, have been deduced for each ion. The MBPT energy results are in excellent agreement with observations, and the mean energy deviation with the NIST observations is 0.05\%. Compared with the elaborate MCDHF/RCI and MCDHF/RCI2 calculations, the accuracy of the MBPT transition data has been estimated to only 1.4\% for transitions among the $n \leq 3$ states for all 13 ions, and 3.0\% for transitions involving the higher states in \ion{Fe}{17}. Because our calculations are systematic and consistent, reporting unified quality of data, we expect that the transition rates are highly accurate and may serve as benchmarks for other calculations.

The present calculations significantly increase the amount of accurate energy data for
a number of Ne-like ions of astrophysics interest, as well as their highly accurate transition rates. A reanalysis of electron temperature and density in the solar or other astrophysical sources using the current extended dataset in high accuracy allows for a more thorough consistency check with the possibility to identify and include new lines of diagnostic value. Through the comparison, we can point out some observations that may have large errors or wrongly assigned, which have been included in Table~\ref{fig.lev.nistlargedifferences}. The high accuracy of the current data may rule out the possibility that wrongly identified lines enter the analysis.

\acknowledgments
The authors express their gratitude to Dr.~MingFeng~Gu for offering guidance in using his FAC code. We acknowledge the support from the National Natural Science Foundation of China (Grant No.~21503066, No.~11504421, and No.~11474034) and the support from the Foundation for the Development of Science and Technology of Chinese Academy of Engineering Physics (Grant No.~2012B0102012). This work is also supported by NSAF under Grant No.~11076009, the Chinese Association of Atomic and Molecular Data,  Chinese National Fusion Project for ITER No. 2015GB117000, and the Swedish Research Council under contract 2015-04842. One of the authors (KW) expresses his gratefully gratitude to the support from the visiting researcher program at the Fudan University.



\clearpage
\bibliographystyle{aasjournal}
\bibliography{ref}

\allauthors

\listofchanges

\section*{Figure and Table}

\clearpage

\clearpage
\clearpage
\clearpage

\clearpage
\begin{deluxetable}{llllllllr}
	\tablecaption{Level energies (in eV) of the states in Ne-like ions from \ion{Cr}{15} to \ion{Kr}{27}, as well as level designations in both the \emph{LSJ}- and \emph{jj} coupling schemes, and the dominant mixing coefficients of the $LSJ$ basis. \label{tab.lev.sub}}
	\tablehead{
		$Z$  &Key       &Conf       &$LSJ$     & \multicolumn{1}{c}{$jj$\tablenotemark{a,b,c}}      &$J^{\pi}$ &\multicolumn{2}{c}{Energy} & \multicolumn{1}{c}{Mixing coefficients}  \\
		\cmidrule(r){7-8}   \cmidrule(r){9-9}
		& & & & & &{NIST\tablenotemark{d}}  &{MBPT\tablenotemark{e}}  & \multicolumn{1}{c}{$LSJ$\tablenotemark{f}} \vspace{ -1.5mm}
	}
	\startdata
  $26$& $  1$& $2s^{2}2p^{6}  $& $^1\!S_{0}$&       $2p+4(0)0~             $& $0^e$&  $0.000000{\rm E+}00$&  $0.000000{\rm E+}00$&  $-1.00 (  1)                          $\\
  $26$& $  2$& $2s^{2}2p^{5}3s$& $^3\!P_{2}$&       $2p+3(3)3~3s+1(1)4     $& $2^o$&  $7.249150{\rm E+}02$&  $7.252443{\rm E+}02$&  $-1.00 (  2)                          $\\
  $26$& $  3$& $2s^{2}2p^{5}3s$& $^1\!P_{1}$&       $2p+3(3)3~3s+1(1)2     $& $1^o$&  $7.267839{\rm E+}02$&  $7.271388{\rm E+}02$&  $-0.74 (  3)  -0.67 (  5)             $\\
  $26$& $  4$& $2s^{2}2p^{5}3s$& $^3\!P_{0}$&       $2p-1(1)1~3s+1(1)0     $& $0^o$&  $7.375376{\rm E+}02$&  $7.378560{\rm E+}02$&  $-1.00 (  4)                          $\\
  $26$& $  5$& $2s^{2}2p^{5}3s$& $^3\!P_{1}$&       $2p-1(1)1~3s+1(1)2     $& $1^o$&  $7.387243{\rm E+}02$&  $7.390537{\rm E+}02$&  $-0.74 (  5)   0.67 (  3)             $\\
  $26$& $  6$& $2s^{2}2p^{5}3p$& $^3\!S_{1}$&       $2p+3(3)3~3p-1(1)2     $& $1^e$&  $7.552334{\rm E+}02$&  $7.554915{\rm E+}02$&  $-0.90 (  6)   0.42 ( 13)             $\\
  $26$& $  7$& $2s^{2}2p^{5}3p$& $^3\!D_{2}$&       $2p+3(3)3~3p-1(1)4     $& $2^e$&  $7.586933{\rm E+}02$&  $7.589928{\rm E+}02$&  $-0.76 (  7)  -0.55 ( 14)             $\\
  $26$& $  8$& $2s^{2}2p^{5}3p$& $^3\!D_{3}$&       $2p+3(3)3~3p+1(3)6     $& $3^e$&  $7.603239{\rm E+}02$&  $7.606096{\rm E+}02$&  $ 1.00 (  8)                          $\\
\enddata
\tablenotetext{a}{The number at the end or inside of the bracket is 2$J$.}
\tablenotetext{b}{$s+ = s_{1/2}$, $p- = p_{1/2}$, $p+ = p_{3/2}$, $d- = d_{3/2}$, $d+ = d_{5/2}$, $f- = f_{5/2}$,  $f+ = f_{7/2}$, $g- = g_{7/2}$,  $g+ = g_{9/2}$, $h- = h_{9/2}$,  and $h+ = h_{11/2}$.}
\tablenotetext{c}{The number after $\pm$ is the occupation number of the corresponding sub-shell. For example, the $jj$ configuration of level 2\\ is $2s_{1/2}^2 2p_{1/2}^2 2p_{3/2}^{3} 3s_{1/2}$.}
\tablenotetext{d}{The observed energies from the NIST ASD \citep{Kramida.2015.V.p}.}
\tablenotetext{e}{The present MBPT results.}
\tablenotetext{f}{The mixing coefficient of the $LSJ$ basis of the state indicated by the key in parenthesis.}
\tablecomments{Only the 201 levels in Ne-like Fe are shown here. Table~\ref{tab.lev.sub} is available online in its entirety in the home page of \emph{ApJS}. \\A portion is shown here for guidance regarding its form and content.}
\end{deluxetable}
\clearpage

\begin{deluxetable}{ccccccc}
	\tabletypesize{\normalsize}
	\tablecolumns{7}
	\tablewidth{0pt}
	\tablecaption{{\textnormal {Wavelengths($\lambda$, in {\AA}), line strengths(\emph{S}, in atomic units), weighted oscillator strengths (\emph{gf}, dimensionless) and transition rates(\emph{A}, in s$^{-1}$) for the transitions in Ne-like ions from \ion{Cr}{15} to \ion{Kr}{27}. }} \label{tab.tr.sub}}
	\tablehead{$Z$  &$j-i$   &Type  &$\lambda$    &$S$  &$gf$  &$A$ \vspace{ -4mm} \\}
	\startdata
$26$& $ 2- 1$& M2 &  $1.7103{\rm E+}01$&  $1.013{\rm E-}01$&  $4.525{\rm E-}08$&  $2.063{\rm E+}05$\\
$26$& $ 3- 1$& E1 &  $1.7059{\rm E+}01$&  $6.862{\rm E-}03$&  $1.222{\rm E-}01$&  $9.335{\rm E+}11$\\
$26$& $ 3- 2$& M1 &  $6.6341{\rm E+}03$&  $1.118{\rm E+}00$&  $6.816{\rm E-}07$&  $3.443{\rm E+}01$\\
$26$& $ 4- 3$& M1 &  $1.1529{\rm E+}03$&  $8.879{\rm E-}01$&  $3.114{\rm E-}06$&  $1.563{\rm E+}04$\\
$26$& $ 5- 1$& E1 &  $1.6784{\rm E+}01$&  $5.518{\rm E-}03$&  $9.986{\rm E-}02$&  $7.882{\rm E+}11$\\
$26$& $ 5- 2$& M1 &  $8.9783{\rm E+}02$&  $1.347{\rm E+}00$&  $6.069{\rm E-}06$&  $1.674{\rm E+}04$\\
$26$& $ 5- 3$& M1 &  $1.0384{\rm E+}03$&  $3.670{\rm E-}01$&  $1.429{\rm E-}06$&  $2.948{\rm E+}03$\\
$26$& $ 5- 4$& M1 &  $1.0448{\rm E+}04$&  $1.089{\rm E+}00$&  $4.214{\rm E-}07$&  $8.582{\rm E+}00$\\
\enddata
	\tablenotetext{\space}{Note: Only transitions among the lowest 27 states of $2s^2 2p^6$ and $2s^2 2p^5 3l$ configurations in \ion{Fe}{17} are shown here. Table~\ref{tab.tr.sub} is available online in its entirety in the home page of \emph{ApJS}. A portion is shown here for guidance regarding its form and content.}
\end{deluxetable}
\clearpage

\begin{deluxetable}{lllcccccc}
	\tablecaption{Comparisons of the experimental and theoretical energies in \ion{Fe}{17}.\label{tab.lev.Fe}}
\tablehead{Key&Config.&LSJ&\multicolumn{6}{c}{Energy}\\
\cmidrule(r){4-9} 
&&&\multicolumn{2}{c}{Exp.}&\multicolumn{4}{c}{Cal.}	\\
\cmidrule(r){4-5}\cmidrule(r){6-9}  
&&&NIST$^a$&CHAINTI$^b$&MBPT$^c$&MCDHF/RCI$^d$&MCDHF/RCI2$^e$&MR-MP$^f$\\
}
\startdata
1  & $2s^{2}2p^{6}  $& $^{1}\!S_{0}$&  0.0000  &  0.000  &   0.0000  &  0.0000  &  0.0000  &  0.0000  \\
2  & $2s^{2}2p^{5}3s$& $^{3}\!P_{2}$&  725.2443&  725.223&   724.9150&  725.1672&  725.1969&  725.170 \\
3  & $2s^{2}2p^{5}3s$& $^{1}\!P_{1}$&  727.1388&  727.138&   726.7839&  727.0773&  727.1015&  727.060 \\
4  & $2s^{2}2p^{5}3s$& $^{3}\!P_{0}$&  737.8560&  737.889&   737.5376&  737.8046&  737.8303&  737.815 \\
5  & $2s^{2}2p^{5}3s$& $^{3}\!P_{1}$&  739.0537&  739.073&   738.7243&  739.0032&  739.0243&  738.997 \\
6  & $2s^{2}2p^{5}3p$& $^{3}\!S_{1}$&  755.4915&  755.485&   755.2334&  755.2505&  755.5067&  755.462 \\
7  & $2s^{2}2p^{5}3p$& $^{3}\!D_{2}$&  758.9928&  758.980&   758.6933&  758.7627&  759.0026&  758.939 \\
8  & $2s^{2}2p^{5}3p$& $^{3}\!D_{3}$&  760.6096&  760.599&   760.3239&  760.3758&  760.6175&  760.564 \\
\enddata
	\tablenotetext{a}{The observed energies from the NIST ASD~\citep{Kramida.2015.V.p}. }
	\tablenotetext{b}{The observed energies from the Chianti database~\citep{{DelZanna.2015.V582.p56,Dere.1997.V125.p149}}.}
	\tablenotetext{c}{The present MBPT energies.}
	\tablenotetext{d}{The present MCDHF/RCI energies.}
	\tablenotetext{e}{The MCDHF/RCI2 energies calculated by~\citet{Jonsson.2014.V100.p1}.}
	\tablenotetext{f}{The MR-MP energies calculated by~\citet{Ishikawa.2009.V79.p25301}.}
\end{deluxetable}
\clearpage

\begin{deluxetable}{lclccc}
	\tabletypesize{\large}
	\tablecolumns{9}
	\tablewidth{0pt}
	\tablecaption{{\textnormal {Level Energies (in eV) for the states which the NIST experimental values differ from the MBPT results by more than 0.2\%. }} \label{tab.lev.nistmbptlardif}}
	\tablehead{\colhead{\emph{Z}} &\colhead{Key\tablenotemark{a}}  &\colhead{State}  & \multicolumn{2}{c}{ Energy } &\colhead{Difference (\%)} \\
		\cmidrule(r){4-5}
		& & &\colhead{MBPT\tablenotemark{b}} &\colhead{NIST\tablenotemark{c}}  \vspace{ -6mm} \\}
	\startdata
$24$&	$29	$&$2s    2p^{6}3s~^1\!S_0$&   $707.2378$&	$712.822  $& $-0.78$\\
$24$&	$87	$&$2s^{2}2p^{5}5d~^3\!P_1$&   $884.1274$&	$886.24   $& $-0.24$\\
$26$&	$29	$&$2s    2p^{6}3s~^1\!S_0$&   $864.8332$&	$869.1    $& $-0.49$\\
$26$&	$57	$&$2s^{2}2p^{5}4d~^1\!D_2$&   $1007.727$&	$1010.682 $& $-0.29$\\
$26$&	$67	$&$2s^{2}2p^{5}4f~^3\!F_4$&   $1015.338$&	$1017.9   $& $-0.25$\\
$26$&	$73	$&$2s^{2}2p^{5}4f~^3\!G_4$&   $1027.657$&	$1014.2   $& $ 1.33$\\ 
$28$&	$28	$&$2s    2p^{6}3s~^3\!S_1$&   $1031.886$&	$1036.26  $& $-0.42$\\
$28$&	$29	$&$2s    2p^{6}3s~^1\!S_0$&   $1038.504$&	$1043.45  $& $-0.47$\\
\enddata
	\tablenotetext{a}{The index number of the level given in Table~\ref{tab.lev.sub}.}
	\tablenotetext{b}{The present MBPT energies.	\tablenotetext{c}{The NIST recommended energies \citep{Kramida.2015.V.p}.}
}
\end{deluxetable}
\clearpage

\begin{deluxetable}{clllllll}
	\tablecaption{Comparisons of the oscillator strengths ($gf$) for the transitions among the $n\leq 3$ levels in \ion{Fe}{17}. The MBPT and MCDHF/RCI \\values, as well as the MCDHF/RCI2 and NIST results, are listed for comparison.\label{tab.tr.n2.n3}}
	\tablehead{
		$j-i$&\multicolumn{1}{c}{Transition}&Type&\multicolumn{4}{c}{$gf$}\\
		\cmidrule(r){4-7} 		
		&&&MBPT\tablenotemark{a}&MCDHF/RCI\tablenotemark{b}&MCDHF/RCI2\tablenotemark{c}&NIST\tablenotemark{d}&\vspace{ -6mm} \\
	}
\startdata
$2 -1 $&   $2s^{2}2p^{5}3s    ~^3\!P_2-2s^{2}2p^{6}      ~^1\!S_0$&  M2&     $4.525{\rm E-}08$&      $4.599{\rm E-}08$&      $4.559{\rm E-}08$&       \nodata         \\
$3 -1 $&   $2s^{2}2p^{5}3s    ~^1\!P_1-2s^{2}2p^{6}      ~^1\!S_0$&  E1&     $1.222{\rm E-}01$&      $1.232{\rm E-}01$&      $1.219{\rm E-}01$&       $1.22{\rm E-}01$\\
$5 -1 $&   $2s^{2}2p^{5}3s    ~^3\!P_1-2s^{2}2p^{6}      ~^1\!S_0$&  E1&     $9.986{\rm E-}02$&      $1.028{\rm E-}01$&      $1.013{\rm E-}01$&       $1.05{\rm E-}01$\\
$7 -1 $&   $2s^{2}2p^{5}3p    ~^3\!D_2-2s^{2}2p^{6}      ~^1\!S_0$&  E2&     $1.019{\rm E-}04$&      $1.023{\rm E-}04$&      $1.023{\rm E-}04$&       \nodata         \\
$10-1 $&   $2s^{2}2p^{5}3p    ~^3\!P_2-2s^{2}2p^{6}      ~^1\!S_0$&  E2&     $1.074{\rm E-}04$&      $1.087{\rm E-}04$&      $1.085{\rm E-}04$&       \nodata         \\
$14-1 $&   $2s^{2}2p^{5}3p    ~^1\!D_2-2s^{2}2p^{6}      ~^1\!S_0$&  E2&     $1.253{\rm E-}04$&      $1.270{\rm E-}04$&      $1.267{\rm E-}04$&       \nodata         \\
$17-1 $&   $2s^{2}2p^{5}3d    ~^3\!P_1-2s^{2}2p^{6}      ~^1\!S_0$&  E1&     $9.718{\rm E-}03$&      $1.018{\rm E-}02$&      $9.864{\rm E-}03$&       $9.70{\rm E-}03$\\
$18-1 $&   $2s^{2}2p^{5}3d    ~^3\!P_2-2s^{2}2p^{6}      ~^1\!S_0$&  M2&     $1.102{\rm E-}06$&      $1.120{\rm E-}06$&      $1.109{\rm E-}06$&       \nodata         \\
\enddata
	\tablenotetext{a}{The present MBPT oscillator strengths.}
	\tablenotetext{b}{The present MCDHF/RCI  oscillator strengths.}
	\tablenotetext{c}{The MCDHF/RCI oscillator strengths given by~\cite{Jonsson.2014.V100.p1}.}
	\tablenotetext{d}{The oscillator strengths recommended by the NIST ASD.}
\end{deluxetable}
\clearpage
\end{document}